# The Equilibrium Vapor Pressures of Ammonia and Oxygen Ices at Outer Solar System Temperatures

B.P. Blakley[a,b,c], Will M. Grundy[c,a], Jordan K. Steckloff [d,e], Sugata P. Tan[d], Jennifer Hanley[c,a], Anna E. Engle[a,c], Stephen C. Tegler[a], Gerrick E. Lindberg[f,g], Shae M. Raposa[a,c], Kendall J. Koga[c,a], Cecilia L. Thieberger[a,c]

a. Department of Astronomy and Planetary Science, Northern Arizona University, 527 S. Beaver St., Flagstaff, AZ, US 86011-6010
b. Pasadena City College, 1570 E. Colorado Blvd., Pasadena, CA, US 91106
c. Lowell Observatory, 1400 W Mars Hill Rd., Flagstaff, AZ, US 86001
d. Planetary Science Institute, 1700 E Fort Lowell Rd. STE 106, Tucson, AZ, US 85719
e. Department of Aerospace Engineering and Engineering Mechanics, University of Texas at Austin, 301 E. Dean Keeton St. C2100, Austin, TX, US 78712-2100
f. Department of Chemistry and Biochemistry, Northern Arizona University, 700 Osborn Dr., Flagstaff, AZ, US 86011
g. Center for Material Interfaces in Research and Applications, Northern Arizona University, 1900 S Knoles Dr. Flagstaff, AZ 86011

# Abstract

Few laboratory studies have investigated the vapor pressures of the volatiles that may be present as ices in the outer solar system; even fewer studies have investigated these species at the temperatures and pressures suitable to the surfaces of icy bodies in the Saturnian and Uranian systems (<100 K, <$10^{-9}$ bar). This study adds to the work of Grundy et al. (2024) in extending the known equilibrium vapor pressures of outer solar system ices through laboratory investigations at very low temperatures. Our experiments with ammonia and oxygen ices provide new thermodynamic models for these species' respective enthalpies of sublimation. We find that ammonia ice, and to a lesser degree oxygen ice, are stable at higher temperatures than extrapolations in previous literature have predicted. Our results show that these ices should be retained over longer periods of time than previous extrapolations would predict, and a greater amount of these solids is required to support observation in exospheres of airless bodies in the outer solar system.

# 1. Introduction

The equilibrium vapor pressure of a substance is the pressure at which the rate of sublimation and the rate of condensation are in equilibrium, and is a function of temperature. Equilibrium vapor pressure is also used to calculate the enthalpy of sublimation, which is the amount of energy needed to change an amount of a substance from solid to gas. Given the temperature and vapor pressure of a certain volatile, it is possible to calculate how long a given amount should be extant on the surface of an icy satellite or dwarf planet depending on its distance from the Sun, or other sources of heating. In other words, a known vapor pressure provides a temperature at which an otherwise volatile solid is stable over a long period of time. Furthermore, accurate predictions of volatile stability can provide testable predictions of which volatiles should be present on the surfaces and in exospheres of airless bodies in the Solar System (e.g., Schaller and Brown 2007).

A 2009 review paper by Fray and Schmitt gathered previous laboratory studies of the vapor pressures of volatile species, and provided thermodynamic models based on these studies to predict respective vapor pressures at any given temperature. Figure 1 shows the results of several of these studies, along with the extrapolations recommended by Fray and Schmitt. Many of these studies were carried out at temperatures higher than 100 K, which are most relevant to the inner Solar System. For certain species, including ammonia, researchers interested in that regime must extrapolate by several orders of magnitude to predict vapor pressures at temperatures relevant to the outer Solar System. The review authors note that these models would be improved by further studies at the more extreme ranges (Fray and Schmitt 2009).

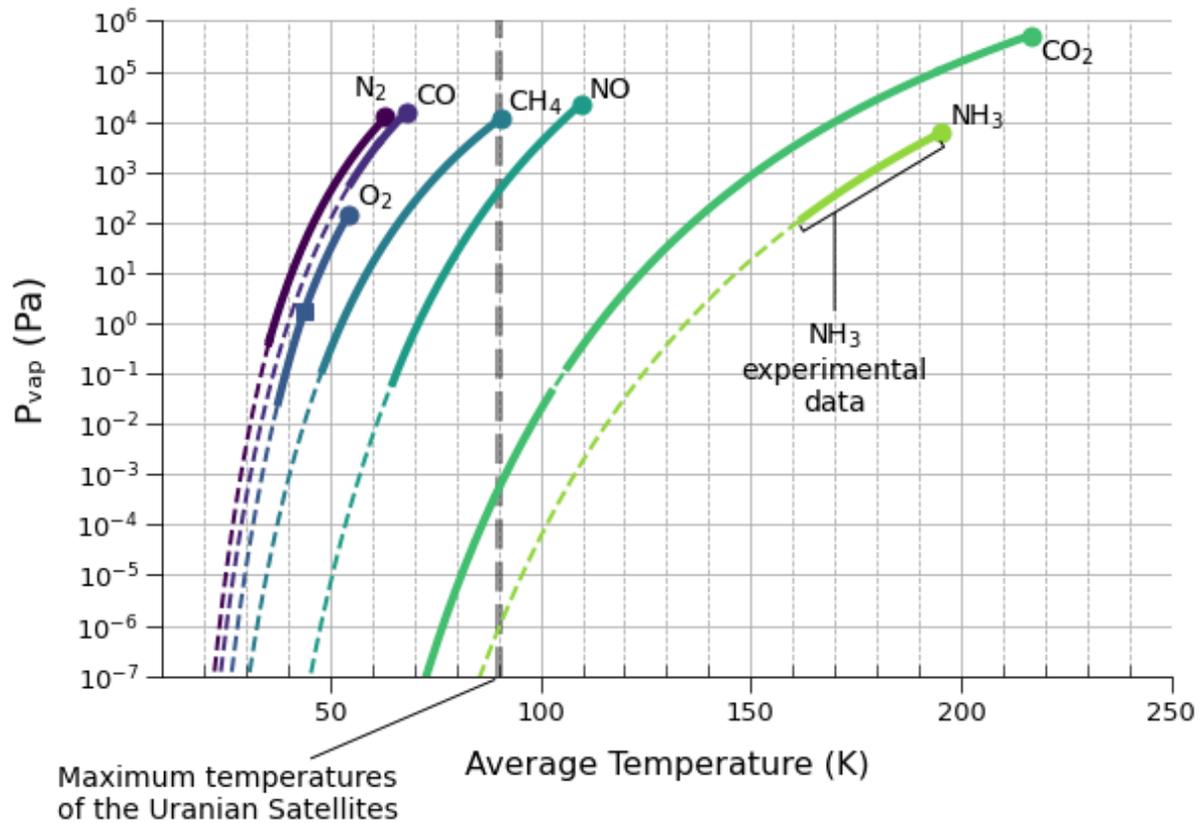

**Fig. 1**. Published laboratory studies and extrapolated values of equilibrium vapor pressures. Adapted from Fray and Schmitt (2009), dotted lines indicate selected extrapolations from published studies; thick lines represent previously published laboratory data; circles show triple points; a square on the $O_2$ line indicates the transition point between beta and gamma crystalline phases (the transition point between alpha and beta is below this range). A thick dotted line at 90 K represents the approximate highest temperature of the Uranian satellites. Many studies have been carried out at higher temperatures than are relevant for the Uranian system, especially ammonia, as annotated in the plot.

Ammonia is very common on icy worlds in the outer solar system. It has been observed on the Uranian moons Ariel and Umbriel, and is predicted to be present on other Uranian satellites (Cartwright et al. 2020, 2023), which have maximum surface temperatures of about 85-90 K (Hanel et al. 1986, Veverka, Brown, and Bell 1991). Ammonia in Pluto's water-ice mantle has been proposed to be essential for lowering the viscosity of water ice, enabling geologic activity on Pluto (Trowbridge et al. 2016, Singer et al. 2022), consistent with the later discovery of ammonia in recently active parts of Pluto's surface (Dalle Ore et al. 2019). This is also consistent with the discovery of ammonia in the icy plumes emanating from Saturn's moon Enceladus (Waite et al. 2009), suggesting a crucial role in cryovolcanism. Ammonia may also contribute to cryovolcanism on Saturn's largest moon Titan. Titan likely formed in an ammonia-methane-water-rich environment that differentiated to include an ammonia-water layer, where the ammonia acts as an anti-freeze (Fortes et. al 2007; Mitri et al. 2008).

Oxygen ice has been observed on the surface of Ganymede (Spencer et al. 1995), Callisto, and Europa (Spencer and Calvin 2002) and was found in unexpectedly high abundance in the coma of comet 67P (Bieler et al. 2015, Mousis et al. 2016, Bodewits and Saki 2022). Furthermore, $O_2$ exospheres have been found across the icy moons of the giant planets, including Saturn's mid-size moons such as Dione, Rhea (Teolis and Waite 2016), and the Galilean Moons of Jupiter Europa (Hall et al. 1995), Callisto (Cunningham et al. 2015), and Ganymede (Hall et al. 1998). Although these oxygen molecules are thought to be radiolytically produced (e.g., Johnson et al. 2004), the exospheric dynamics of these species likely see them cycling between solid and vapor phases (e.g., Mogan et al. 2020; Steckloff et al. 2022). As such, understanding the vapor pressure behavior of these species at low temperatures (and thus their energetics) is crucial to understanding and modeling these moons' exospheres.

Oxygen ice has three crystalline phases: alpha ($\alpha$) stable below 23.78 K, beta ($\beta$) between 23.78 and 43.77 K, and gamma ($\gamma$) above 43.77 K (Prokhvatilov, Galtsov, and Raenko 2001). As shown in Figure 1, results from Hoge (1950) and Aoyama and Kanda (1934) provide 13 experimental data points in total on the vapor pressures of $\beta$ and $\gamma$ oxygen ice at temperatures ranging from 37.6–53.09 K and at pressures from $2.7 \times 10^{-7}$ to $1.01 \times 10^{-3}$ bar (Fray and Schmitt 2009). Studies of ammonia ice were carried out at higher temperatures, with 15 experimental data points from Karwat (1924) and Overstreet and Giauque (1937) ranging from 162.39–195.36 K with pressures from $1.27 \times 10^{-3}$ to $6.08 \times 10^{-2}$ bar (Fray and Schmitt 2009), also plotted in Figure 1.

In a study published in 2024, Grundy et al. showed that previous extrapolations of the vapor pressure of CO, $N_2$, and $CH_4$ ices at very low temperatures predicted higher sublimation rates than were found in laboratory investigation. This was not entirely unexpected, as the previous temperatures and pressures studied required extrapolation over several orders of magnitude. Nonetheless, these results have implications for the stability of these species on icy bodies; in particular, these solids should be retained on airless bodies for longer periods than previously expected (Grundy et al. 2024). Given the results of this study, it is reasonable to conclude that studies of other outer Solar System ices may show similar results, especially those with extrapolations over multiple orders of magnitude, as shown in Figure 1. Moreover, we are interested in confirming and extending the literature on outer solar system materials through additional laboratory study (e.g., Raposa et al. 2022). Results from these studies show that refinement of extrapolated values through laboratory investigation is both useful and worthwhile.

Over long periods of time, small discrepancies in vapor pressures and enthalpies of sublimation can cause a considerable difference in the amount of ices retained. All other variables being equal, the rate of loss into a vacuum is proportional to the vapor pressure, so the length of time before a quantity of material has disappeared is inversely proportional to the vapor pressure. Therefore, in this study, we extend the previous studies by examining the vapor pressures of $O_2$ and $NH_3$ at temperatures relevant for these bodies.

## 2. Methods

To explore the vapor-solid equilibrium (VLE) of each species at very low temperatures, we employ an experimental technique that uses a quartz crystal microbalance (QCM), which is sensitive enough to measure the mass difference of sublimation and condensation of gas molecules onto or off of a <1-$\mu$m-thick ice film (e.g., Sack and Baragiola, 1993; Allodi et al., 2013; Luna et al., 2014, 2018; Hudson et al., 2022). Our experiments are completed within a stainless steel vacuum chamber at the Astrophysical Materials Laboratory at Northern Arizona University (NAU), as described in Grundy et al. (2024). A turbopump on the chamber provides a background pressure of 1.0 to 2.0 x $10^{-8}$ torr; with the attached helium cryocooler running, pressures down to 3.0 to 4.0 x $10^{-9}$ torr are achieved (see Grundy et al. 2024). The second stage of the cryocooler is shielded, but the QCM is not covered by the shield. Also attached to the chamber are two variable leak valves, or input valves, by which we add gas species at controlled pressures, which then condense onto the QCM. In each experimental run, we add gas species to the chamber to deposit a thickness of ice. After deposition is complete, we close all valves and increase the temperature of the chamber, causing the species to sublime, allowing us to measure sublimation rates.

Throughout the experiment, we record the QCM frequency, temperature, and pressure in 3-second intervals. By measuring the frequency change (due to deposition onto or sublimation off of the QCM) over a time interval at a constant temperature, we can compute the change in areal mass density, d$Q$/d$t$ (Grundy et al. 2024, eq. 1). With d$Q$/d$t$, we can then calculate the effective gas pressure at the QCM, which we denote as $p_{QCM}$ (Grundy et al. 2024, eq. 3). The pressure gauge ($p_{gauge}$) is mounted within the chamber but with no line of sight to the QCM and cannot measure the pressure directly on the surface of the ice; by calculating $p_{QCM}$ and measuring $p_{gauge}$, we can obtain a correction factor, $\Phi$, which is equal to $p_{QCM}$ divided by $p_{gauge}$ (Grundy et al. 2024). Finally, we can compute the equilibrium vapor pressure ($p_{vap}$) by solving the Hertz-Knudsen-Langmuir equation (Langmuir 1913), shown below, utilizing the calculated values above (Grundy et al. 2024).

$$\frac{dQ}{dt} = -p_{vap}\sqrt{\frac{M}{2\pi R T_{QCM}}}$$

During the set up and course of the experiment, we monitor and record the temperature of the cold head, pressure of the chamber, the frequency of the QCM, and utilize a mass spectrometer to monitor relative levels of the species being studied as well as any contaminant species. See Grundy et al. (2024) for details of the recording and control devices.

As stated above, we begin each experiment by depositing a gas species; once we have ice deposited on the QCM, we begin the sublimation stage. We will first describe the deposition process, followed by the sublimation process.

## 2.1 Deposition

Before every experiment, the chamber was heated externally for a minimum of 16 hours, or for multiple days when changing between volatile species, to bake out contaminant gasses. At the start of an experiment, after turning off the external heaters, the initial background pressure was noted, as measured by the internal pressure gauge. We then turned on the cryocooler and set an initial deposition temperature using the temperature controller. We began monitoring and recording the temperature, pressure, and QCM frequency at this time. The initial temperature was chosen based on the volatile species being studied; it must be a temperature well below the expected lower bound of sublimation, at which the ice will be stable during the full deposition phase. In order to determine the deposition temperature, we made an initial estimate based on known freezing temperatures of the species, and then ran an initial test to check for stability. For $O_2$ ice we set the deposition temperature to 10 K, and to 70 K for $NH_3$.

When the temperature of the system had stabilized at the determined deposition temperature, we began stepped deposition. Noting the background pressure inside the cooled chamber, we opened the input valve to let gas into the chamber until the pressure gauge read 2-3 times the starting pressure. Keeping the pressure stable, we tracked the change in QCM frequency over a period of 3-5 minutes, allowing enough time for the frequency to reach a constant rate of change. We then increased the pressure 2-3 times the previous pressure, allowed the system to stabilize, and tracked the QCM frequency again. We repeated this process, recording up to ten steps of pressure increase, until we reached a chamber pressure on the order of magnitude of $10^{-4}$. This stepped process allowed us to calculate the correction factor, $\Phi$, as stated above.

We measured deposition thickness by the change in frequency of the QCM in Hz. Figure 2 shows the recorded data from the QCM monitor, pressure monitor, and temperature monitor over the course of a typical deposition period. Both plots show elapsed time along the x-axis; the top plot tracks the QCM frequency in Hz, and the middle plot tracks the pressure in Torr. The temperature of the cold head is kept stable over the deposition period. At time zero, we began adding gas to the chamber, which can be observed by the increase in pressure at that time in the middle plot; the stepped deposition as described above can also be observed in this plot throughout the deposition period. As we increase the pressure of the gas in the chamber, gas condenses onto the QCM at a higher rate, slowing the frequency of the QCM, shown in the top plot. A decrease in QCM frequency corresponds to an increase in deposition thickness. Typically, we deposited enough material to cause a decrease in QCM frequency of 2.0 to 3.0 kHz, which translates to approximately 0.1-0.2 μm onto the QCM, but when measuring sublimation rates at higher temperatures, we deposited enough material to cause a decrease in frequency of up to 10.0 kHz. Once the desired thickness was reached, we closed the input valve and allowed the pressure inside the chamber to stabilize; when we closed the input valve, this caused a sudden drop in the pressure of the chamber, as is visible in the middle plot of Figure 2 at 0.73 hours. When we stop adding gas to the chamber, deposition onto the QCM also ends; at that time the frequency of the QCM stabilizes to a constant value.

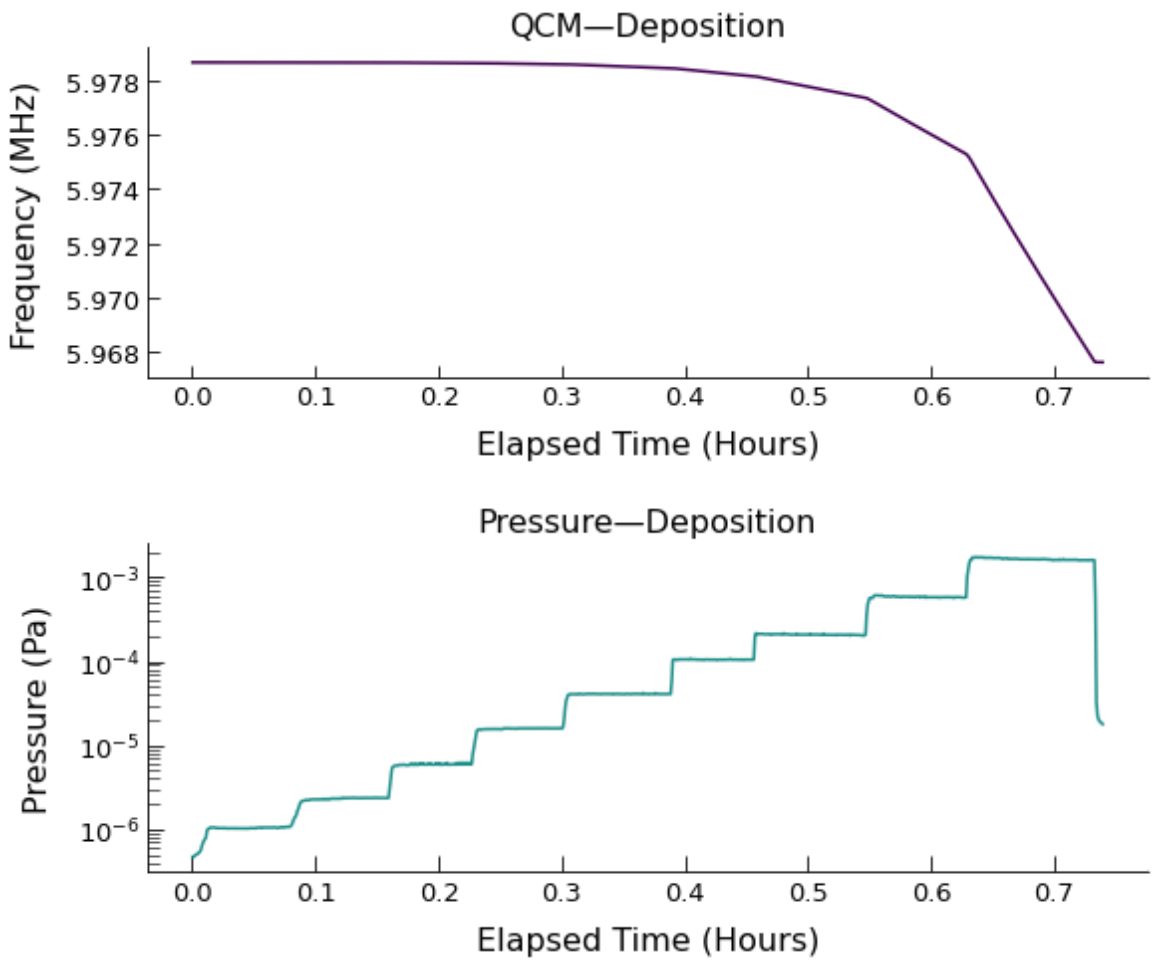

**Fig. 2.** Stepped deposition process. Plots show the frequency of QCM (top) and chamber pressure (bottom) during the deposition process of one experiment. Note the decreasing frequency of the QCM corresponding to the increase of pressure; as pressure increases in the chamber, more gas is deposited onto the QCM, and the thickness of the ice on the QCM increases at a faster rate.

## 2.2 Sublimation

Sublimation measurements begin after the deposition period is completed, and as soon as the pressure has stabilized; the chamber is a closed system during the entirety of the sublimation period. During an initial sublimation experiment for each species of gas, we would first find the lower bound of sublimation before continuing with further sublimation measurements. The lower bound of sublimation is the lowest temperature at which the volatile will sublime off of the QCM at a steady, measurable rate in this pressure regime. In order to characterize the equilibrium vapor pressure of these substances in the outer solar system, where temperatures can be much lower than 100 K, we must make laboratory measurements at temperatures as low as possible. Finding the minimum sublimation temperature possible in our vacuum chamber for each volatile species allows us to gather data points at temperatures relevant to airless bodies such as Pluto and the Uranian satellites. These measurements can then be used to refine the curves extrapolated from measurements at higher temperatures by previous researchers, which are shown in Figure 1.

At the beginning of the initial sublimation experiment, as well as each thereafter, the cold head would still be set to the deposition temperature. We started the initial sublimation experiment by increasing the temperature of the chamber from the deposition temperature to an estimated lower bound, given previous literature; for example, 30 K for oxygen. Once the temperature stabilized, we would watch the QCM for a steady increase in frequency (which denotes decreasing mass on the QCM), and the pressure monitor for a rise in pressure. At 30 K for oxygen, we saw a modestly increasing slope of about 10 Hz/min on the QCM while the temperature was stable. Similarly, at 90 K for ammonia, the QCM frequency increased by ~2 Hz/min. In both cases, this provided an *approximate* lower bound of sublimation.

Figure 3 shows an entire sublimation experiment, beginning after deposition, including the tests for upper and lower bounds. The first dashed vertical line marks the first increase in pressure that corresponds to a stable temperature and a modestly increasing QCM frequency. After finding this approximate lower bound, we would then increase the temperature to find an approximate upper bound. This was achieved by stepping up the temperature, first in increments of five degrees and allowing the QCM to stabilize, until the stable sublimation rate was over 10 Hz/min, and then by two degrees, until the sublimation rate was over 100 Hz/min. This temperature provided an *approximate* upper bound. During this period of stepped increases, we also gathered measurements on the net mass loss rate from the QCM at each temperature step.

After finding the approximate lower and upper bounds, we would determine a true lower bound by decreasing down to the approximate lower bound and then decreasing another 1-3 K until the rate of change of QCM frequency was < 1 Hz/min but still measurable and steady. This provided the *true* lower bound. From that point, we gathered sublimation data, i.e., the net mass loss rate at stepped temperature increases, until reaching the approximate upper bound. Finally, we continued in 1-2 K increments until the sample was lost from the QCM. This temperature provided our *true* upper bound.

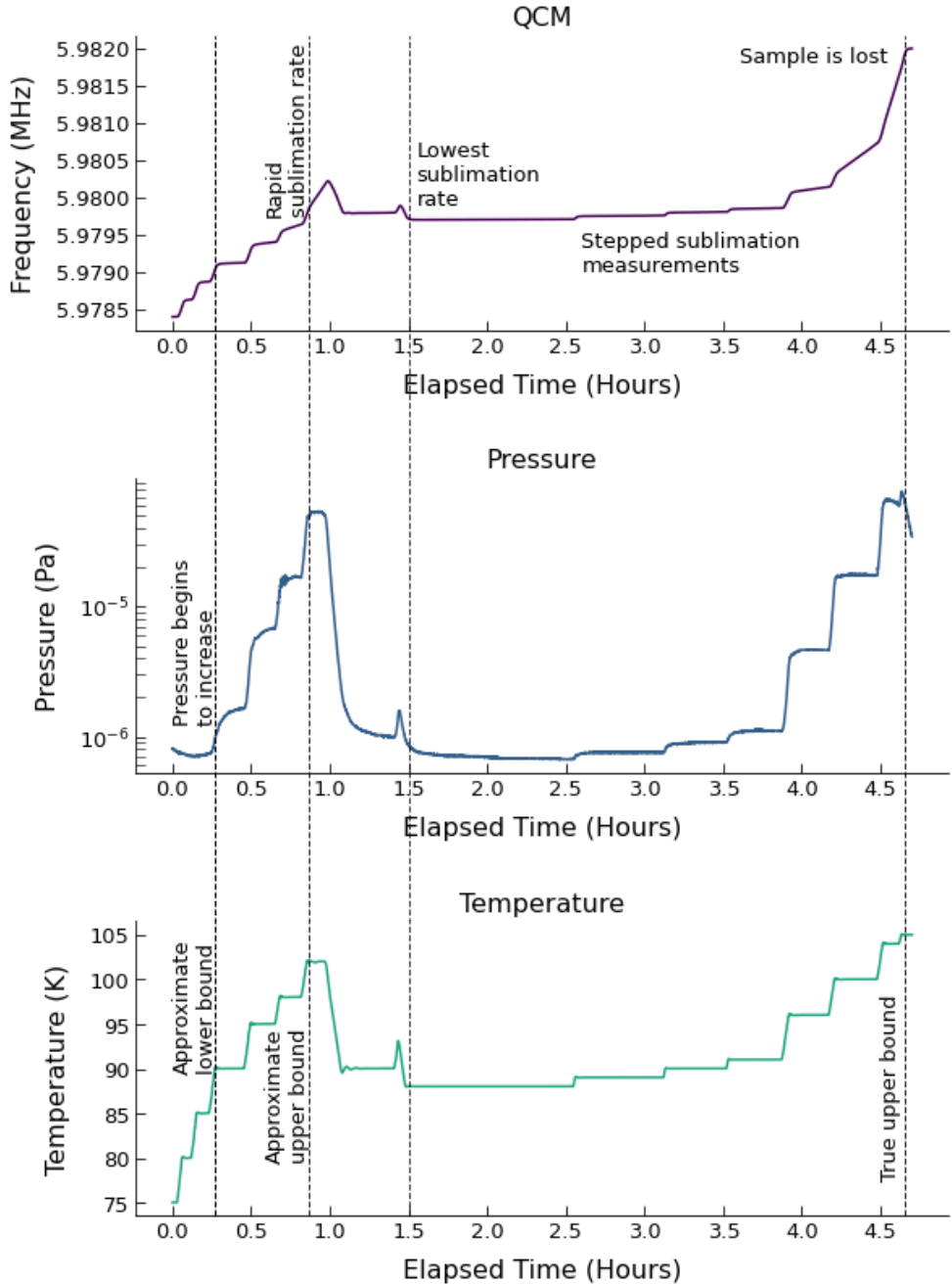

**Fig. 3.** Stepped sublimation measurements with lower and upper bounds. Plots show QCM frequency, chamber pressure, and QCM temperature, respectively. First dotted line indicates when the first (modest) increase in QCM frequency was seen, providing the approximate lower bound. Second dotted line indicates the actual lower bound, during which <1 Hz/minute increase was seen on the QCM. Final dotted line indicates the upper temperature where the remaining sample was lost via sublimation from the QCM.

These two bounds would then be utilized as our window for further measurements. During subsequent sublimation experiments, we gathered pressure and QCM data stepwise at each whole number temperature. For each temperature step, we recorded a start time and end time during which the temperature and pressure was relatively stable. A minimum of three minutes was recorded after temperature stabilization at each step, however the amount of time needed to measure sublimation rate at a given temperature is dependent on the rate, which is lower at lower temperatures (therefore a longer time at that temperature is required). For the lowest temperatures, at least an hour of pressure and frequency data was recorded (see the lowest sublimation rate in Figure 3).

Some redeposition onto the QCM by the species under study means that we only measured the net mass loss rate, not the sublimation mass loss rate that could be directly related to the vapor pressure. In order to reduce this effect, after ice deposition, the background pressure was reduced by first increasing the temperature at the cold head to 5-10 K above the lower bound, and then decreasing steadily to the target temperature. The cold head was the coldest point in the chamber, so material would freeze onto it and adjacent parts. During this warming period, other surfaces in the chamber were able to warm; some of the ice would sublime from them and be removed by the turbopump. Some would remain, and we accounted for the deposition of this material onto the QCM based on the chamber pressure and $\Phi$ correction that we measured earlier, but minimizing the background gas helps reduce the deposition term and thus the uncertainty it adds to the result.

In any vacuum system, some amount of background contaminants will be present. Contaminants were reduced via baking the chamber overnight prior to an experiment with the turbopump running continuously. Generally, the substance being measured has a much higher partial pressure in the chamber (e.g., $>10^{-11}$ bar) than all other background contaminants combined, so their effect is negligible. But at the lowest pressures (e.g., near $10^{-12}$ bar), the influence of the background contaminants starts to become significant. We gathered sublimation data at temperatures and pressures below our lower bound; this allowed us to determine the point where background contaminants begin to have an effect on the measurements by noting when the data begins to bend away from the trend when plotted logarithmically. Any measurements at that point or lower were not included in our calculations. Finally, we note the larger degrees of uncertainty with larger error bars along the pressure axis at the lowest points.

## 3. Calculation

We utilize the data gathered during the deposition periods of each experimental run to calculate temperature and pressure corrections for the corresponding sublimation data. With the correction factors from the deposition periods, we can then calculate the pressure at the surface of the QCM ($p_{QCM}$) at each sublimation step. Figure 4 illustrates the analysis process for one such step. We then use the calculated $p_{QCM}$ for each temperature step to calculate the vapor pressure. Finally, we can derive the enthalpy of sublimation.

### 3.1 Temperature and Pressure Correction

We have two silicon diode thermometers in the chamber, one on the cold head of the assembly, and other on the holder of the QCM. They are nearly the same temperature, but we average the two temperatures together to provide the nearest estimate of the temperature directly on the surface of the QCM. Then, a linear correction model is applied. The corrected temperature is equal to constant $a$ multiplied by the averaged temperature, plus constant $b$, where $a = 0.996$, and $b = 0.290$, based on measurement of two phase changes, a solid-solid phase change in $CH_4$ ice at 20.4 K, and the melting of $C_3H_8$ ice at 85.5 K.

We also calculate the correction factor of the pressure monitors, $\Phi$. Other methods (e.g., Luna et al., 2014, 2018) may not include this step, however this correction factor allows us to account for condensation of gas molecules from elsewhere in the chamber; the addition of $\Phi$ results in a slight increase in the calculated vapor pressures, mitigating the effect of background pressure. First, we must obtain a starting frequency, ending frequency, a starting time, and ending time with a stable, decreasing slope on the QCM, during multiple periods of stable pressure and temperature, which we gather during the stepped deposition period. These allow us to calculate the change in areal density over time ($dQ/dt$), which further allows us to solve for the pressure on the QCM ($p_{QCM}$), with given constants such as the temperature of the room, and the mass in g/mol of the gas species (see Grundy et al. 2024, Basic Equations). We solve for $\Phi$ by dividing $p_{QCM}$ over the pressure measured on the pressure gauge ($p_{gauge}$), and then fitting a model to these points (Grundy et al. 2024). This pressure correction factor is then used in the analysis of the sublimation data and is calculated for each run of the experiment.

## 3.2 Equilibrium Vapor Pressure

We calculate the vapor pressure from net sublimation data gathered at temperature steps, which is reduced in a similar manner to the deposition steps. Utilizing the start and end times recorded during the experiment for each temperature step, we plot the QCM, pressure, and temperature over that time period. Figure 4 shows an example of one full sublimation step, with the QCM frequency, pressure, and temperature over the same time period. We refine the data by cutting the start and end times in each step to the most stable portion. We calculate the average pressure and temperature during each period of stability and fit a line to the QCM frequency to find the slope. Using the Φ correction factor calculated from the deposition steps, we also calculate $p_{QCM}$ at each temperature by multiplying $p_{gauge}$ by Φ. From this, we calculate $dQ/dt$. Finally, we can calculate the equilibrium vapor pressure ($p_{vap}$) for each recorded temperature as stated in eq. 6 in Grundy et al. (2024):

$$p_{vap} = \Phi\, p_{gauge} \sqrt{\frac{T_{QCM}}{T_{room}}} - \frac{dQ}{dt}\sqrt{\frac{2\pi R T_{QCM}}{M}},$$

where $T_{QCM}$ is the corrected temperature as described in section 3.1, $R$ is the universal gas constant, and $M$ is the molar mass of the volatile species.

We apply an expression given by Lobo and Ferreira (2001) to provide a thermodynamic model that relates the equilibrium vapor pressure $p_{vap}$ (in bar) to an exponential dependence on a function of temperature:

$$ln(p_{vap}) = A - B/T + C\, ln(T) + D_2 T + D_3 T^2 + D_4 T^3$$

where the parameters $A$, $B$, $C$, $D_2$, $D_3$, and $D_4$ are each derivable analytically using properties of the material. The expressions for the parameters in terms of material properties are given in detail in Lobo and Ferreira (2001). In this work, whereas constants $C$, $D_2$, $D_3$, and $D_4$ are derived from the isobaric heat capacities of solid phase and ideal gas, $A$ and $B$ are fitted over experimental vapor-pressure data as in Grundy et al. (2024). For beta $O_2$, we fit $A$ and $B$ over our new experimental data along with the beta $O_2$ data from Aoyama and Kanda (1934); for $NH_3$, we fit $A$ and $B$ over our new experimental data in addition to the data from Karwat (1924), Burrell and Robertson (1915), and Overstreet and Giauque (1937). Heat capacities of solid are taken from Davies (1956) for $NH_3$ and Giauque and Johnston (1929) for beta $O_2$, while that of ideal gas is 4 (i.e., 8/2) $R$ for $NH_3$, and 7/2 $R$ for beta $O_2$. The values of these parameters are listed in Table 1 and the data and fits are shown in Figures 5 and 7.

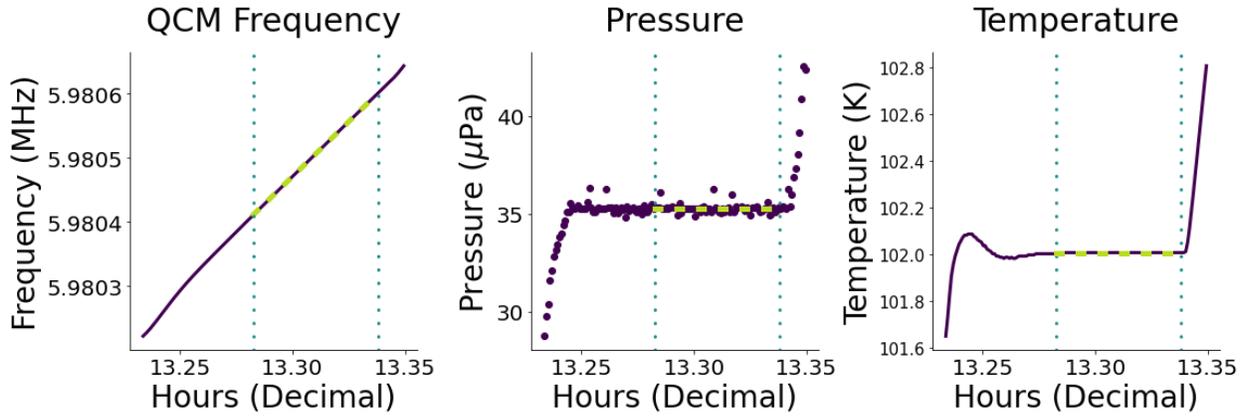

**Fig. 4.** Example of one temperature step of sublimation data collection (in this case, for ammonia). Blue dotted lines show the bounds where the pressure and temperature are both stable within the system, while the QCM has a stable increasing slope. Thicker, light gray dotted lines show the lines which have been fit to each data point. Sublimation rates are calculated from stable periods like this within each sublimation step.

### 3.3 Enthalpy of Sublimation

We derive the enthalpy of sublimation ($L$) from the above fitted equation, and arrive at a fourth-degree function of temperature as in Lobo and Ferreira (2001):

$$L(T) = R(B + CT + D_2 T^2 + D_3 T^3 + D_4 T^4) \pm \sigma_B R$$

where $\sigma_B$ is the uncertainty from fitting parameter $B$ over the data of sublimation pressure. The enthalpy is plotted against temperature in Figures 6 and 8 for $NH_3$ and beta $O_2$, respectively.

# 4. Results & Discussion

For both ammonia and oxygen ices, we find that experimental vapor pressures are lower than extrapolated values. Consequently, enthalpies of sublimation are higher. Below, we compare our laboratory data to the extrapolations given in Fray and Schmitt (2009), and our derived enthalpies of sublimation to NIST values.

## 4.1 Ammonia

As of today, there is no vapor-pressure experimental data in the literature for ammonia at temperatures as low as ours (Gao et al., 2023). Therefore, for comparison purposes, we use extrapolation calculated by Fray and Schmitt (2009). We see lower vapor pressures for ammonia than the extrapolation by an average factor of 1.5. It is not unexpected that we see a difference in the theoretical value and laboratory value, given that our temperatures are much lower, and our pressures are orders of magnitude lower than the previous laboratory studies that the Fray and Schmitt calculation is based on. Figure 5 shows our data points with uncertainties, and our fitted curve, compared to the extrapolation. Calculating enthalpy of sublimation as a function of temperature, as shown in Figure 6, we find that over the range of 80-120 K, our experimental values are higher by 2.74% on average than the NIST accepted constant value of 31.2 kJ/mol over a temperature range of 177–195 K based on the data measured by Overstreet and Giauque (1937). This work expands on that of Overstreet and Giauque (1937), extending laboratory-measured values of the enthalpy of sublimation of ammonia to lower temperatures, without suggesting any change to the value at the previously measured temperature range.

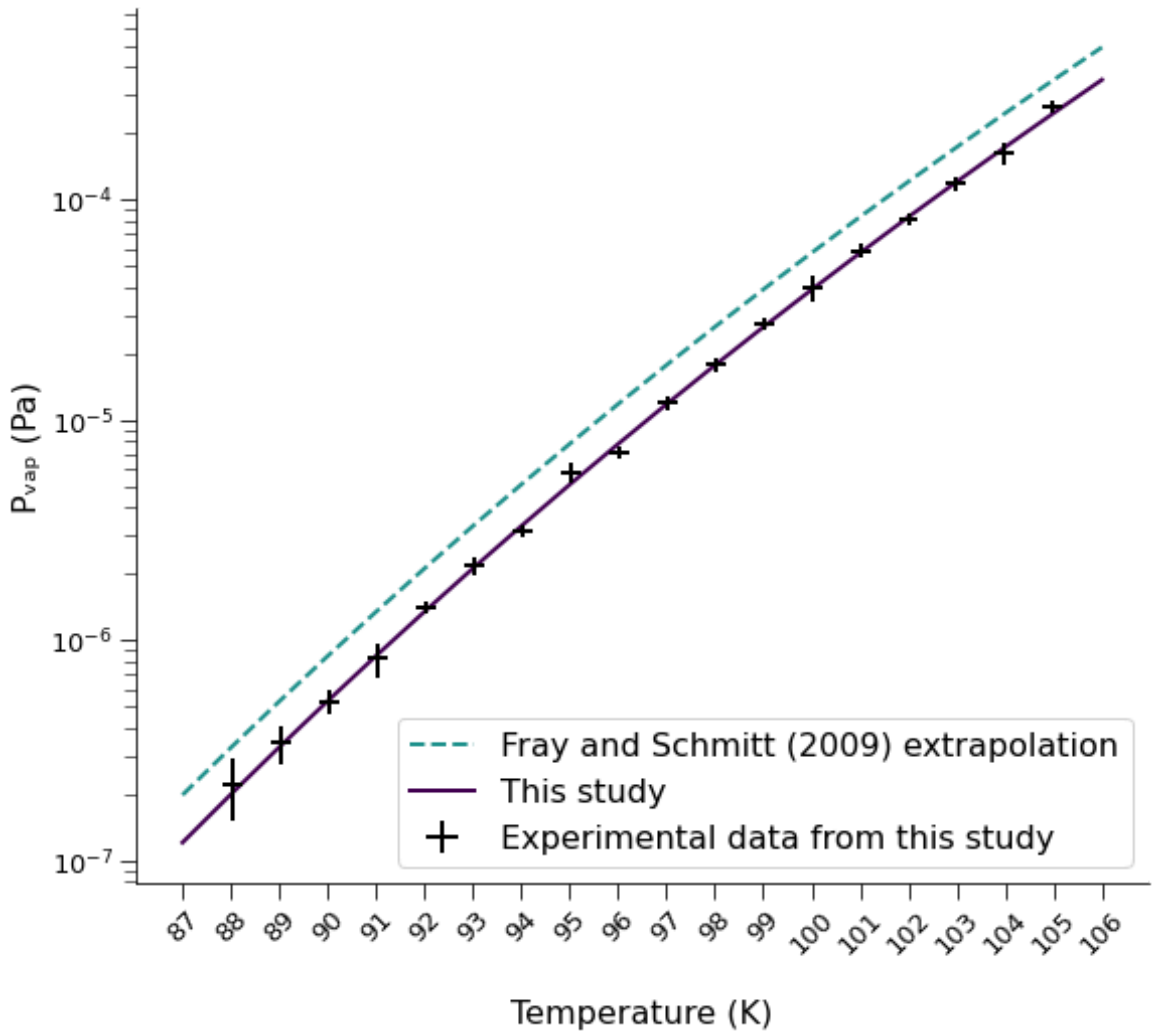

**Fig. 5.** Laboratory vs. extrapolated vapor pressures of ammonia ice. The measured vapor pressures of ammonia ice from this work are shown with bars of uncertainty (black crosses) compared to polynomial extrapolations from Fray and Schmitt (2009)/Brown and Zeigler (1980) (dotted line). The polynomial recommended by this work is shown as a purple solid line. This work shows that vapor pressures are lower than accepted values by a factor of 1.5.

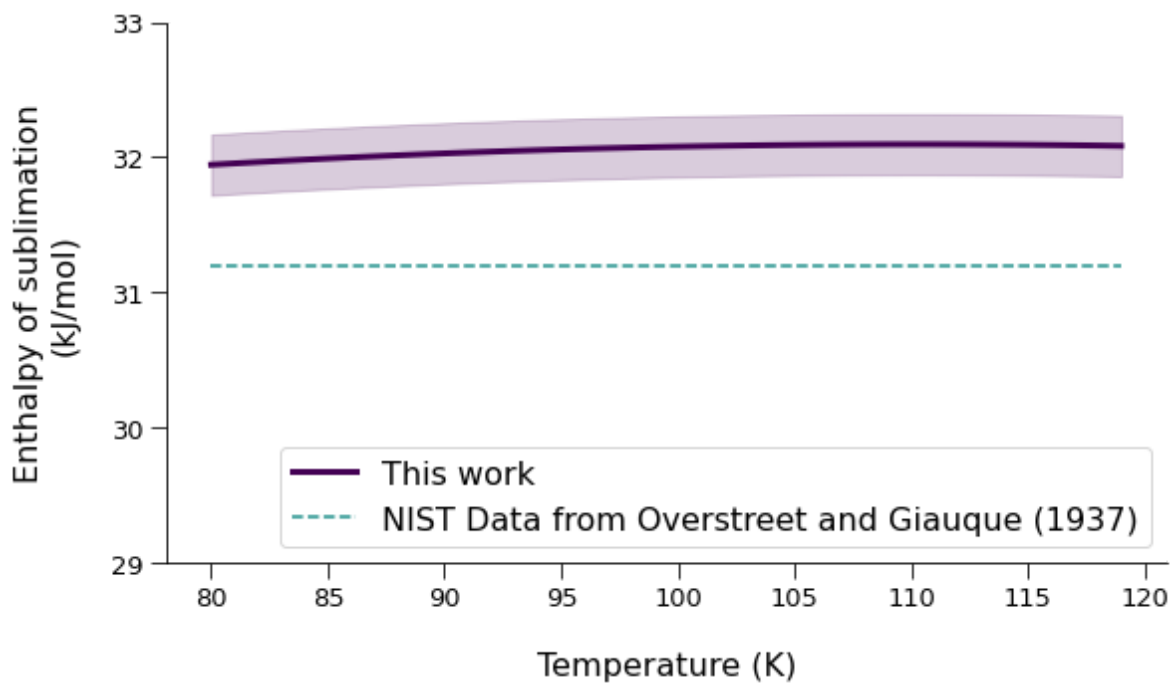

**Fig. 6.** Calculated enthalpy of sublimation of ammonia ice, as a function of temperature. This work is shown as a solid line with a shaded region of uncertainty, compared to the accepted NIST constant value based on data by Overstreet and Giauque (1937) in a temperature range of 177-195 K (dashed line). The values calculated from this study show that the enthalpy of sublimation of ammonia ice over this temperature range is on average 2.7% higher than the values in a higher temperature range as a rough comparison due to the absence of literature data in our temperature range.

## 4.2 Oxygen

For beta oxygen ice, we see lower vapor pressures than previous work by an average factor of 1.6. Our data and fitted curve are compared to previous literature in Figure 7. Again, we calculate the enthalpy of sublimation for oxygen ice as a function of temperature. Over the range of 20-50 K, our calculated values are higher by an average of 9.2% than the NIST accepted constant value of 9.26 kJ/mol (Shakeel et al. 2018), as shown in Figure 8. Our equilibrium vapor pressure data agree well with the vapor pressure data of Shakeel et al. (2018), however we use a different way of calculating the enthalpy of sublimation.

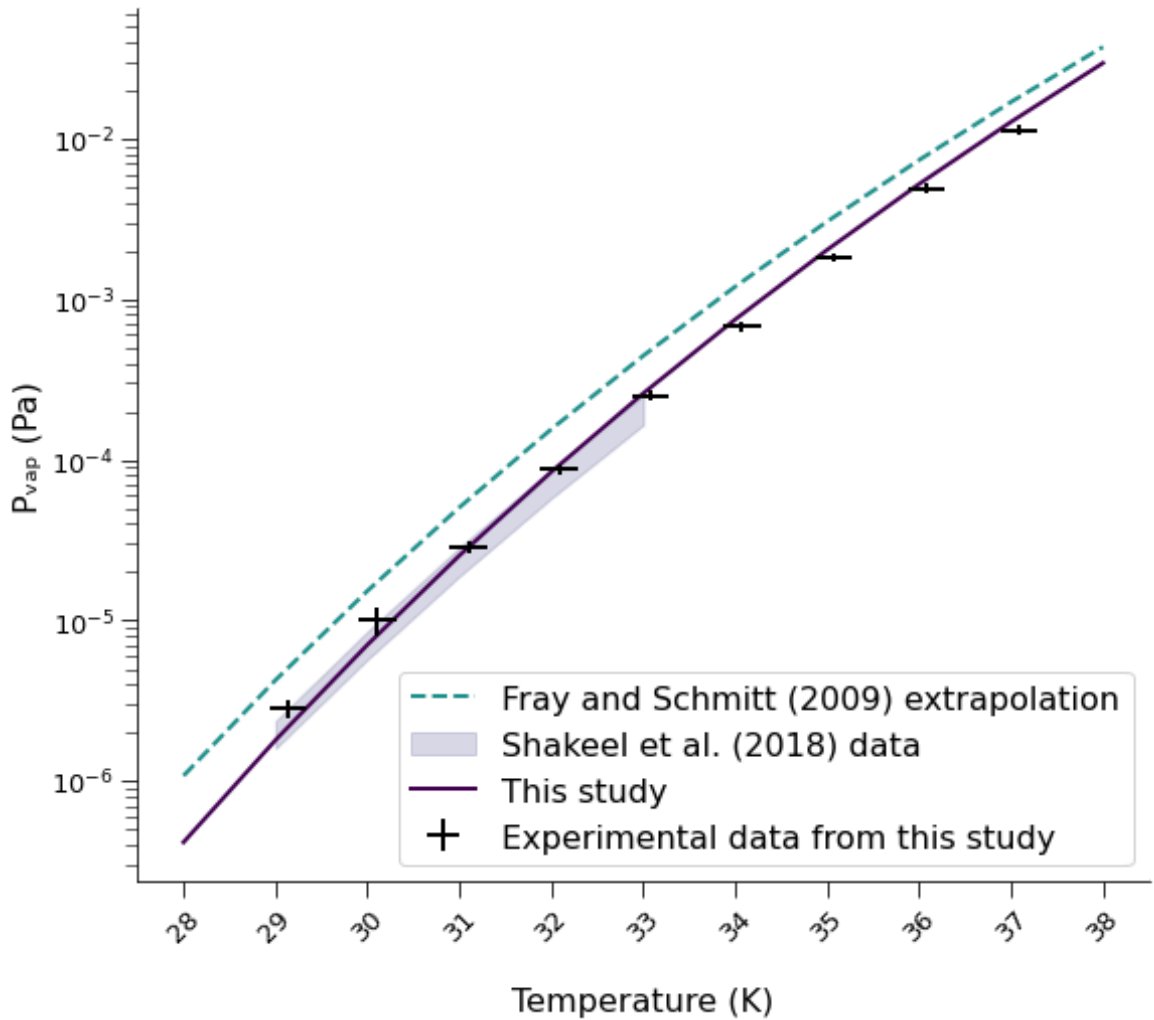

**Fig. 7.** Laboratory vs. extrapolated vapor pressures of oxygen ice. The measured vapor pressures of beta oxygen ice (crystalline β) from this work are shown with bars of uncertainty (black crosses) compared to polynomial extrapolations from Fray and Schmitt (2009)/Brown and Zeigler (1980) (dotted line), and experimental data from Shakeel et al. (2018) (shaded region). The polynomial recommended by this work is shown as a purple solid line. Beta oxygen ice vapor pressures are also lower than the polynomial extrapolation by a factor of 1.6, however they align well with the previous experimental data.

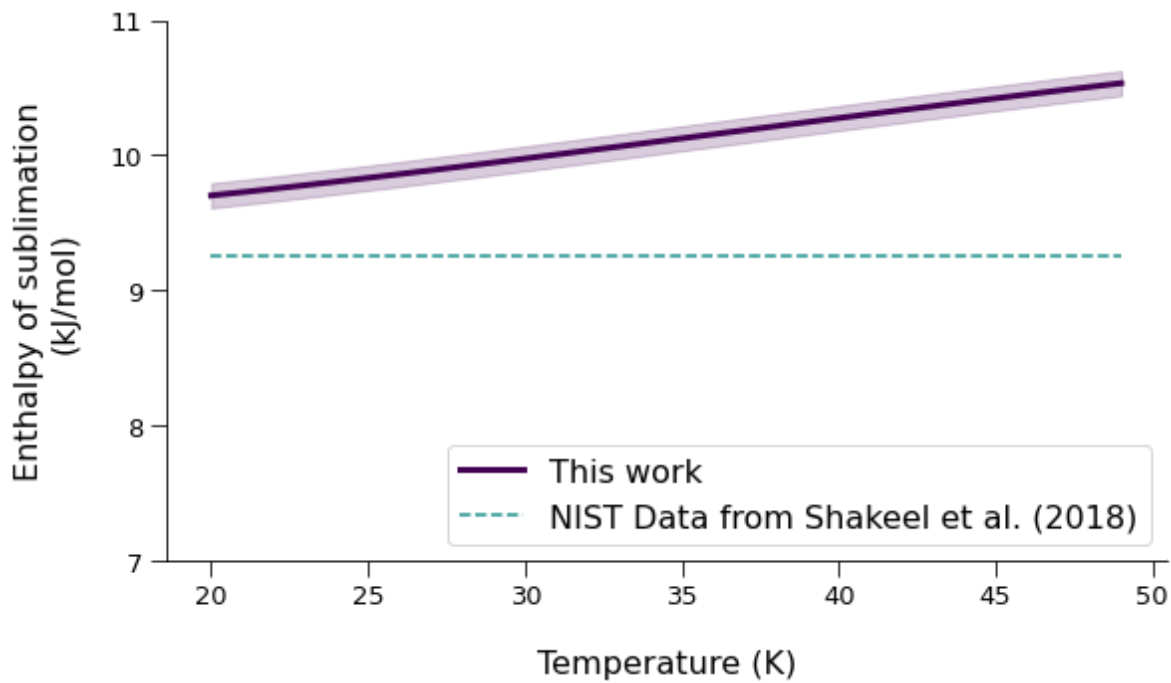

**Fig. 8.** Calculated enthalpy of sublimation of beta oxygen ice as a function of temperature. This work is depicted by a solid line with a shaded region of uncertainty, compared to the accepted NIST constant value from Shakeel et al. (2018) (dashed line). This work shows that the enthalpy of sublimation of beta oxygen ice over this temperature range is on average 9.2% higher than accepted values.

**Table 1**

Fitted and derived constants for the sublimation pressure and the enthalpy of sublimation of each studied species.

| Species | A | B (K) | C | $D_2$ (K$^{-1}$) | $D_3$ (K$^{-2}$) | $D_4$ (K$^{-3}$) |
|---|---|---|---|---|---|---|
| Ammonia | -5.55 ± 0.45 | 3605 ± 31 | 4.82792 | -0.024895 | $2.1669 \times 10^{-5}$ | $-2.3575 \times 10^{-8}$ |
| Beta Oxygen | 15.29 ± 0.29 | 1166.2 ± 8.6 | -0.75587 | 0.14188 | $-1.8665 \times 10^{-3}$ | $7.582 \times 10^{-6}$ |

## 4.3 Sources of uncertainty

In our calculations of the vapor pressure on the QCM, we assume a sticking coefficient of unity, to account for gas molecules that may be arriving at the surface of the ice from elsewhere in the apparatus. Other researchers (e.g., Schmitt and Rodriguez 2003, Haynes et al. 1992) have found that sticking coefficients for less volatile species increase linearly, up to unity, with decreasing temperatures. At the low temperatures of our studies, it is reasonable to assume a sticking coefficient of unity, as in our initial calculations, however we have also included a calculation of the equilibrium vapor pressures of both gasses, if the sticking coefficient was near zero, shown below in Figure 9. With a lower sticking coefficient, the values we calculate for equilibrium vapor pressure would be even lower.

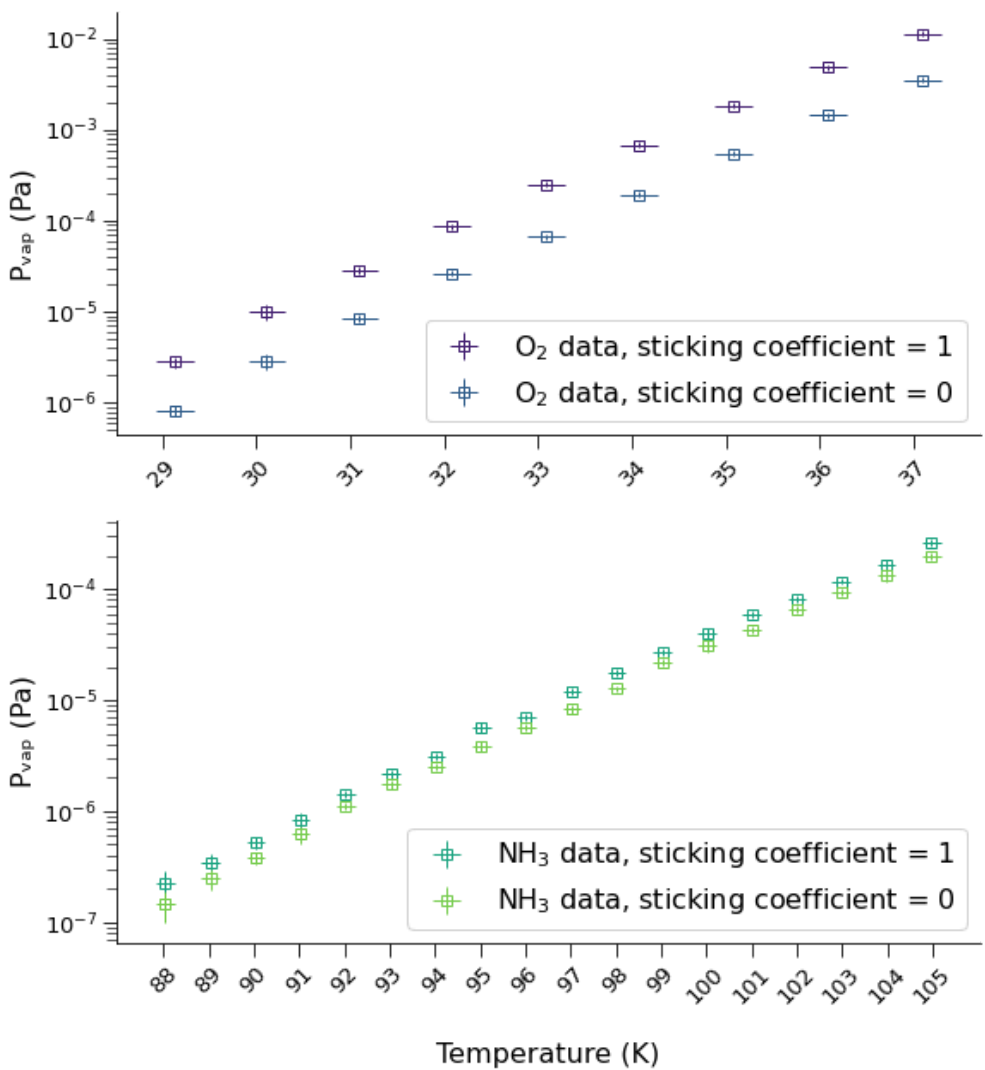

**Fig. 9.** Equilibrium vapor pressures are lower if the sticking coefficient (percent of gas molecules from elsewhere in the chamber impinging onto the ice film that stick to the film) is lower. We assume a sticking coefficient of unity in calculating our results (as shown in Figures 5 and 7), however if the sticking coefficient is lower, the equilibrium vapor pressures of the species must also be lower. The true vapor pressures must be within this range.

We estimate an uncertainty of temperature measurements of ± 0.2 K, following Grundy et al. (2024). Any thermal gradient within the ice caused by the enthalpy of sublimation or external heating is captured by this value for temperature uncertainty. As an example, assuming that the surface of the ice film is being heated by 296 K black body radiation from the chamber walls (a maximum, since the room-temperature stainless steel walls should have less than unit emissivity, and the ice absorbs weakly at thermal infrared wavelengths), and using the lowest conductivity value for the solids within our temperature range (e.g., ~0.5 W/m K at the beta-gamma transition for oxygen ice; Jeżowski et al. 1993), we find the upper limit of the thermal gradient caused by external heating of a 1-$\mu$m film for either species would be < 0.001 K. Any thermal flux from sublimation is more than an order of magnitude lower than that from the limited external heating, so any thermal gradient from the enthalpy of sublimation is negligible.

## 5. Conclusions

We find that the equilibrium vapor pressures studied herein deviate significantly than would be predicted using the curves of Fray and Schmitt (2009) and are lower than these curves would predict. As a corollary, the enthalpies of sublimation are higher than would be expected from extrapolating empirical fits in the literature to low temperatures, especially for oxygen, and for ammonia to a lesser degree. Our models suggest that at even lower temperatures, this trend would continue, with volatiles having higher stability than currently inferred.

These results have important implications for the behavior of these volatile species at relevant temperatures in the outer solar system. Lower equilibrium vapor pressures and, equivalently, higher enthalpies of sublimation imply that less vapor will be present for any fixed area of exposed material. Thus, were $O_2$ to be detected in the exosphere of an outer solar system icy body, more solid $O_2$ would necessarily need to be present on the object's surface longer than previously expected. For example, to produce the detected densities of $O_2$ in the exospheres of Dione, Rhea (Teolis and Waite 2016), Europa (Hall et al. 1995), Callisto (Cunningham et al. 2015), and Ganymede (Hall et al. 1998) would likely require greater exposures of $O_2$ on the surfaces of these moons. Furthermore, the abundances of $O_2$ detected in the coma of 67P (Bieler et al. 2015, Mousis et al. 2016, Bodewits and Saki 2022) would necessarily require greater quantities of solid $O_2$ within the nucleus, assuming that the detected $O_2$ is a result of direct sublimation from solid $O_2$, rather than a daughter product of various reactions within the coma. It is important to note that $O_2$ is unlikely to be found as a pure solid on these surfaces and, instead, is usually trapped within other ices such as $H_2O$ or $CO_2$. Nonetheless, the stability of oxygen solids is likely to affect the stability of binary or ternary ice mixtures involving $O_2$, and these results will be important for researchers utilizing the vapor pressures of pure ices to model the sublimation of such mixtures.

Furthermore, the higher enthalpies of sublimation and lower vapor pressures of these species would likely render their involvement in cryovolcanic activity harder to detect in the vapor phase, yet easier to detect in the solid phase. For example, the ammonia detected on the surface of Pluto and correlated with cryovolcanic activity (Dalle Ore et al. 2019) was likely rendered easier by ammonia's lower-than-expected volatility, as less of the substance would have been lost to the Plutonian atmosphere since formation. Conversely, the detected ammonia in the plumes of Enceladus (Waite et al. 2009) would have required greater quantities of ammonia to produce a detectable signature.

# Data availability

Data files associated with this paper have been archived for public access at https://openknowledge.nau.edu/id/eprint/6254.

# Funding

We acknowledge support from the National Science Foundation Research Experience for Undergraduates Program at Northern Arizona University (NSF Award #1950901). Portions of this work and laboratory facility were supported by NASA Solar System Workings Program grant 80NSSC19K0556; by the State of Arizona Technology and Research Initiative Fund (TRIF), administered by the Arizona Board of Regents; and by philanthropic donations from the John and Maureen Hendricks Foundation and from Lowell Observatory's Slipher Society.

microbalance. *Monthly Notices of the Royal Astronomical Society, 473*(2), pp.1967-1976. https://doi.org/10.1093/mnras/stx2473

Mitri, G., Showman, A.P., Lunine, J.I. and Lopes, R.M., 2008. Resurfacing of Titan by ammonia-water cryomagma. *Icarus, 196*(1), pp.216-224. https://doi.org/10.1016/j.icarus.2008.02.024

Mogan, S.R.C., Tucker, O.J., Johnson, R.E., Sreenivasan, K.R. and Kumar, S., 2020. The influence of collisions and thermal escape in Callisto's atmosphere. *Icarus, 352*, p.113932. https://doi.org/10.1016/j.icarus.2020.113932

Mousis, O., Ronnet, T., Brugger, B., Ozgurel, O., Pauzat, F., Ellinger, Y., Maggiolo, R., Wurz, P., Vernazza, P., Lunine, J.I. and Luspay-Kuti, A., 2016. Origin of molecular oxygen in comet 67P/Churyumov–Gerasimenko. *The Astrophysical Journal Letters, 823*(2), p.L41. https://doi.org/10.3847/2041-8205/823/2/L41

Overstreet, R. and Giauque, W.F., 1937. Ammonia. The heat capacity and vapor pressure of solid and liquid. Heat of vaporization. The entropy values from thermal and spectroscopic data. *Journal of the American Chemical Society, 59*(2), pp.254-259. https://doi.org/10.1021/ja01281a008

Prokhvatilov, A.I., Galtsov, N.N. and Raenko, A.V., 2001. X-ray studies of phase transitions in solid oxygen. *Low Temperature Physics, 27*(5), pp.391-396. https://doi.org/10.1063/1.1374726

Raposa, S.M., Tan, S.P., Grundy, W.M., Lindberg, G.E., Hanley, J., Steckloff, J.K., Tegler, S.C., Engle, A.E. and Thieberger, C.L., 2022. Non-isoplethic measurement on the solid–liquid–vapor equilibrium of binary mixtures at cryogenic temperatures. *The Journal of Chemical Physics, 157*(6). https://doi.org/10.1063/5.0097465
27